# Observation of Giant Spin Splitting and d-wave Spin Texture in Room Temperature Altermagnet $RuO_2$

Zihan Lin[1,†], Dong Chen[2,3,†], Wenlong Lu[1,†], Xin Liang[1], Shiyu Feng[1], Kohei Yamagami[4], Jacek Osiecki[5], Mats Leandersson[5], Balasubramanian Thiagarajan[5], Junwei Liu[6], Claudia Felser[3], Junzhang Ma[1,7,*]

[1]*Department of Physics, City University of Hong Kong, Kowloon, Hong Kong, China*
[2]*College of Physics, Qingdao University, Qingdao 266071, China*
[3]*Max Planck Institute for Chemical Physics of Solids, 01187 Dresden, Germany*
[4]*Japan Synchrotron Radiation Research Institute, 1-1-1, Sayo-cho, Sayo-gun, Hyogo 679–5198, Japan*
[5]*MAX IV laboratory, Fotongatan 8, 22484 Lund, Sweden*
[6]*Department of Physics, The Hong Kong University of Science and Technology, Hong Kong, China.*
[7]*City University of Hong Kong Shenzhen Research Institute, Shenzhen, China*

*†The authors contributed equally to this work.*
*\*Corresponding to: Junzhang Ma (junzhama@cityu.edu.hk)*

**Recently, a new magnetic phase called altermagnetism has been proposed, ushering in a third distinct magnetic phase beyond ferromagnetism and antiferromagnetism. It is expected that this groundbreaking phase exhibits unique physical properties such as C-paired spin-valley locking, anomalous Hall effect, nontrivial Berry phase, and giant magnetoresistance. Among all the predicted candidates, several room temperature altermagnets are suggested to host significant potential applications. Nevertheless, direct evidence about the spin pattern of the room temperature altermagnet is still unrevealed. $RuO_2$ is identified as the most promising candidate for room temperature d-wave altermagnetism exhibiting a substantial spin splitting of up to 1.4 eV in previous research. In this study, utilizing angle-resolved photoemission spectroscopy (ARPES), we report experimental observation of the giant spin splitting in $RuO_2$. Furthermore, employing spin-ARPES, we directly observed the d-wave spin pattern. Our results unequivocally show that $RuO_2$ is a perfect d-wave altermagnet with great potential for upcoming spintronic applications.**

Magnetism, a foundational domain in condensed matter physics, conventionally classifies magnetic materials into two primary phases: ferromagnets and antiferromagnets (although unconventional forms like ferrimagnetism and chiralmagnetism also recognized). Ferromagnets exhibit macroscopic magnetization in spatial space and spin splitting in reciprocal momentum space (Fig.1A). Conversely, typical antiferromagnets show no net macroscopic magnetization in spatial space and no spin splitting in reciprocal momentum space when time reversal symmetry combined with inversion or translation operation is preserved (Fig.1B) (*1–3*). Recent theoretical investigations have identified materials with substantial spin-split band structures but no net macroscopic magnetization (Fig. 1C) (*2–18*). These materials challenge easy classification as either ferromagnets or antiferromagnets. This has led to the proposal of altermagnetism, a third magnetic phase that appropriately describes such materials (*2, 3*). In contrast to conventional antiferromagnets, which have spin sublattices that are linked to one another through straightforward translation operations, altermagnetism involves linking sublattices with opposite spins using rotation operations (*2, 3, 19*). Applying the same rotational angle in momentum space causes to a spin flip between up and down. As outcome of this phenomenon, even-order wave symmetries of the spin pattern, such as the d-, g-, i-waves, and others (*2–4, 20*).

The s-wave Cooper pairs in conventional superconductivitors find analogous counterparts in ferromagnetism. Meanwhile, the investigation of unconventional d-wave superconductivity has been going on for more than two decades, which has led scientists to ask if there is an equivalent unconventional d-wave counterpart in magnetic materials (*21*). Efforts have been made to investigate d-wave magnetism within the context of strong electronic correlations (*22–24*). On the other hand, at the fundamental level of an effective single-particle description of magnetism, altermagnetism presents the prospect of unconventional d-wave or high even-parity wave magnetism (*2–16, 18, 25*). This suggests that altermagnets can inherently possess robust d-wave properties (*2, 3*), which would broaden the category of spin quantum phases.

The investigation of altermagnetism greatly expands the symmetry category in the field of magnetism. Due to their unique physical properties – such as stability under perturbing magnetic fields, strong noncollinear spin currents, anomalous Hall effect, C-paired spin-valley locking, nontrivial Berry phase, giant magnetoresistance - altermagnets appear to be promising candidates for promoting the next generation of information technology, (*5, 8, 11–17, 25–32*). Despite these encouraging prospects, it's important to note that the experimental exploration of the electronic structures of altermagnets is still in its early stages.

Calculations have predicted dozens of potential altermagnet candidates (*2, 3*), and recent research has reported band splitting in materials with gaps of about 0.3 eV in $MnTe_2$ and MnTe, about 0.8 eV for CrSb (*2, 3, 33–35*). Energy splitting and critical temperature are two criteria that are important for the candidates' potential of application. To be applied in spintronics, the critical temperature must be higher than room temperature; also, to increase transport robustness, the energy splitting should be three to five times larger than the room temperature induced energy broadening, that is to see greater than 300-400 meV. Remarkably, of these candidates, only three ($RuO_2$, MnTe, and CrSb) are in the promising region for potential application, as shown in Fig.1E (*3*). However, till now, there hasn't been any reporting of the direct experimental observation of the spin pattern of these intriguing candidates. $RuO_2$ stands out among the promising candidates, exhibiting the largest spin-splitting at around 1.4 eV and high critical temperature more than room temperature (*3, 6, 36–38*). This substantial spin splitting in $RuO_2$ is of great scientific interest and has a wide range of potential uses. Novel observations in $RuO_2$ include the reporting of anomalous Hall effect and in-plane Hall effect (*39–42*), along with the prediction of giant and tunneling magnetoresistance (*43–45*), the identification of titled spin currents (*46*), and the documentation of terahertz emission induced by spin splitting effect (*47, 48*). Additionally, superconductivity has been observed in thin-film $RuO_2$ (*49–51*). Topological superconductors, such as p-wave superconductivity, are suggested to arise from the introduction of superconductivity into altermagnet, where the spin degeneracy is lifted and spin is consistent between $k$ and $-k$ (*52, 53*). Evidence of time reversal symmetry breaking has been reported through the difference between circular plus and minus light-induced photoemission spectra (*54*). Moreover, the presence of nodal lines in the electronic structure has also been studied (*55*).

Despite the several novel findings in $RuO_2$, direct spectroscopic evidence proving the giant spin splitting and d-wave spin pattern has been lacking until now, but they are essential for confirming $RuO_2$ as an ideal altermagnet with substantial application potential. In the current study, we bridge this gap by employing synchrotron-based ultraviolet (UV) ARPES, soft X-ray ARPES, and spin-ARPES. Using these advanced techniques, we provide strong direct evidence that giant spin splitting exists in $RuO_2$, the most optimal room remperature altermagnet candidate. Moreover, our study directly captures the d-wave spin pattern in the momentum space, which represents a significant advance in the experimental verification of $RuO_2$'s unique altermagnetic properties.

The crystal structure of $RuO_2$, shown in the inset of Fig. 1D, hosts the space group denoted as P42/mnm (SG136). The Ru atom is surrounded by six O atoms, forming an octahedron with 2-fold rotation symmetry. The spin configuration of Ru within the two sublattices can be connected through a 90-degree rotation operation combined with a translation operation, which breaks parity time symmetry. The crystal and spin configuration distinctly characterize the d-wave altermagnetism that is anticipated in $RuO_2$. X-ray photoemission spectroscopy shows both O and Ru peaks in Fig. 1D. Fig. 1F and G, respectively, show the band structures in the paramagnetic phase and altermagnetic phase. Double-degenerate bands are shown in the paramagnetic phase, while in the altermagnetic phase, distinct spin splitting is visible along the ΓM and AZ directions. In the graphical representation, the red color bands indicate spin-up states, whereas the blue color bands indicate spin-down states. The significant spin-splitting along the ΓM direction is especially noteworthy. Our predicted calculation shows energy splitting as high as 1.54 eV as indicated by the double arrows along ΓM direction in Fig.1G, which is similar to the previously reported 1.4 eV (*3*). This makes it the most substantial among the candidates proposed in the recent predictions (Fig.1E).

Breaking of parity-time symmetry results in band degeneracy lifting of altermagnet with respect to the paramagnetic phase. Experimentally, the observation of band splitting provides strong evidence for the establishment of long-range magnetic order. The specific cut along the ΓM direction provides an excellent point of reference for verifying this claim. Both UV and soft X-ray photon energy-dependent mappings were conducted in order to precisely determine the correct photon energies and locations in the momentum space, as shown in Fig. 2B and C. The natural cleavage of the sample along the (110) direction, causes the central point to alter between Γ and M with varying photon energy. The soft X-ray ARPES data in Fig. 2C distinctly reveals a clear periodic structure. In order to define the photon energies associated with high symmetry planes that align well with the lattice parameters and set the inner potential, vertical cuts at the center of the Brillouin Zone (BZ), as shown in Fig. 2D, offer insights into the band structure along the ΓM direction (see more details in Supplementary Materials section 1 and Fig. S1). On the other hand, owing to the $k_z$ broadening effect, the out-of-plane dispersion in UV ARPES data is weaker but still noticeable, as shown in Fig. 2B. Since the inner potential remains consistent for both soft X-ray and UV measurements, we double confirmed the $k_z$ relationship with photon energy by deriving the UV photon energies associated to high symmetry planes from the soft X-ray data. This derivation, as shown in Fig. S1C, exhibits excellent agreement with the UV data. To investigate the in-plane Fermi surface, two photon energies, 115 eV and 74 eV, were chosen to measure the electronic structure associated with the Γ and M planes, respectively, as plotted in Fig. 2E and F.

We designate the direction parallel to ΓM as $k_x$ and the direction parallel to ΓZ as $k_y$. Plots of the horizontal cuts crossing the BZ center are shown in Figs. 2G and H, which both match to the ΓM spectra but have different central locations. Within these ARPES spectra, we identify four bands near Fermi level labeled as α - δ. The α band has a flat characteristic that almost exactly on the Fermi level. The adjacent β and γ bands cross the Fermi level between Γ and M. Along the perpendicular ΓZ direction, these two bands degenerate, as illustrated in Fig. 2E, where we also identify surface states (SS) forming a quasi-1-D Fermi surface along with one small pocket. The δ band is broad in energy and it bends back at the binding energy of approximately 0.4 eV. This band dispersion is consistent with the results of DFT calculations, as shown in Fig. 2J. Soft X-ray Fermi surface maps at the Γ plane is presented in Fig. 2K, without the presence of surface bands. The cut along the ΓM direction is plotted in Fig. 2J, with the DFT calculation superimposed. Apart from the flat α band at the Fermi level, bands β to δ are clearly identified. The double-peak in momentum distribution curve

(MDC) in Fig. 2I clearly shows the separation of the β and γ bands. To verify if the α band is a surface state, a surface band structure projection calculation is given in Fig. S2. The results show no discernible evidence of surface bands along the ΓM direction on the Fermi level near α band, but one dispersive hole surface band located at around -0.25 eV above the δ band (see more details in Supplementary Materials section 2). In a previous study, a flat surface band was linked to the bulk nodal line (*55*). However, even if such surface states exist, our computational analysis reveals that they are topologically unprotected and overlap with the bulk band projection, basically extending the wavefunction of the same bulk flat band onto the top layer. This observation suggests that bulk spectra are included in the α band. We explain the vanishing of the α band at soft X-ray area by a slight chemical potential shift between the surface and bulk regions caused by the band bending effects near the surface. Upon detailed comparison, we found that a calculated band structure with energy shift upward of 0.36 eV agrees well with the UV-APRES data. Meanwhile, a 0.41 eV upward shift of the calculated bands is necessary to align with the soft X-ray ARPES data, where the flat α band is just above the Fermi level. On the other hand, detailed analysis about the polarization dependent data (see Supplementary Materials section 2, and Fig. S3) shows that along ΓM direction, one surface band is strengthened under LV polarization while bulk bands are apparent under LH polarization. Circular polarization, on the other hand, has both surface and bulk states along the same cut.

In the paramagnetic phase, the electron bands crossing the Fermi level ($E_F$) and centered around Γ (*β* and *γ* bands) are initially double degenerate. The double degeneracy does not lift in the whole BZ as illustrated in Fig. S4. On the other hand, this degeneracy is lifted when the long-range magnetic order begins to emerge. In Fig. 2G,H,I, we experimentally observed band splitting between *β* and *γ* bands near the $E_F$ along the ΓM direction. It indicates the formation of long-range magnetic order and provides evidence for the requirement of altermagnetism in $RuO_2$.

In the calculations presented in Fig. 1G, considerable spin splitting is shown in ΓMAZ plane. We investigate the band structure obtained at of 115 eV photon energy to examine the band splitting within this plane. The corresponding Fermi surface and 3D electronic structure are shown in Fig. 3C and D. Figures 3E-J plot experimental results for a set of bands parallel to the ΓM direction, which are designated as cut 1-6. While Fig. 3K-P show calculated results for the same bands. Along the ΓM direction, the spin-polarized α band, which is located exactly on the Fermi level, is most isolated in energy space close to the middle of the ΓM. The Kramer pair band with opposite spin is located at around 1 eV above the Fermi level. Together, the δ band and the α band share the same spin orientation, forming a large, purely spin-polarized energy range spanning from -0.7 eV till above $E_F$. This property renders $RuO_2$ an ideal and robust candidate for spintronic applications. Experimental data in Fig. 3E-J show detailed band splitting along cut 2-6, which agree well with the calculations (Fig. 3K-P). Meanwhile, the occupied band exhibits the largest spin splitting up to 1.54 eV in the energy range of -0.7 to -2.5 eV. Fig. S6 shows the experimental data for a wide energy range, revealing the signature of this huge spin splitting. Additionally, Fig. S7 presents a comparison between computations and experimental results along cut 1-6 collected with soft X-rays. The general band dispersion in the experimental data is in agreement with that obtained with UV light, except the bands are broader and the energy resolution is lower. This comprehensive comparison provides solid evidence that the experimental observations and theoretical predictions under the altermagnetic phase are consistent.

Although the band splitting characterized the altermagnetic phase, direct confirmation of the d-wave altermagnetism in $RuO_2$ requires the measurement of spin-resolved band structure and the associated d-wave spin pattern. To achieve this, we conducted spin-ARPES measurements using a VLEED detector connected to the DA30 Scienta Omicron detector. Given the unique characteristic of a d-wave spin pattern, wherein the spin direction undergoes a flip after a 90-degree rotation, we carefully chose four points symmetrically around the M point along the ΓM direction (Fig. 4A-C) in the out-of-plane momentum space within the ΓMX plane, as indicated in Fig.4R, where the spin-splitting is prominent in the calculation (Fig. 4B,C). The spin-up and spin-down energy distribution curves (EDCs) For the left *k* point are plotted in Fig. 4F and the zoomed-in view is shown in Fig. 4G. It is shown that the density of states for spin-up is larger than that for spin-down between -0.7 eV and the Fermi level. Fig.4H displays the intensity difference between spin up and down, indicating a spin-up polarization for the $\alpha$ and $\delta$ bands that is in good agreement with the prediction as depicted in Fig. 4B (EDC1). Meanwhile, the right point exhibits similar results, as seen in Fig. 4O,P,Q. Furthermore, we varied the photon energy to tune the momentum space to the lower points (54 eV) and upper point (95 eV). Fig. 4I,J and K present the corresponding EDCs for spin-up and spin-down of the lower point. As expected, the density of states for spin-down is larger than that for spin-up from -0.7 eV up to the Fermi level. Concurrently, the upper point exhibits a comparable spin polarization, as seen in Fig. 4L,M,N, indicating the spin down polarization. Fig. 4D displays the spin polarization of the four EDCs, which reaches a maximum of around 10%. Notably, the EDCs exhibit a considerable background, leading us to mitigate its impact. To address this, we applied a fitting procedure to the four EDCs, including a linear background and Gaussian peaks multiplied by Fermi Dirac function, as demonstrated in Fig. S8. For both spin-up and spin-down EDCs, the same linear background is subtracted at the same location in order to eliminate the impact of background on the spin differential. The spin polarization is greatly increased by this background subtraction, with the largest polarization up to 20% at both the upper and lower points. By aggregating the spin polarizations from all four points, we obtained compelling and direct evidence supporting the d-wave spin pattern of altermagnetism in $RuO_2$.

In conclusion, we have successfully conducted a systematic study of $RuO_2$ by employing synchrotron-based UV ARPES, soft X-ray ARPES, and spin ARPES. Not only have we observed the spin-splitting, but we have also directly witnessed the d-wave spin pattern in the ideal altermagnetic material, $RuO_2$. As emphasized earlier, altermagnets exhibit great scientific promise and have the potential for diverse applications. Meanwhile, the lack of experimental investigations into the spin-polarized electronic structures of altermagnet $RuO_2$ underscores the significance of our findings. As the experimentally validated ideal candidate with large spin splitting and with critical temperature much higher than room temperature, $RuO_2$ not only exhibits novel physical properties but also presents itself as a promising candidate for the development of next-generation spintronic devices in the near future (*40, 41, 43–45*).

**Acknowledgements**

We acknowledge Prof. Yaobo Huang, Mr. Gexing Qu for help during the spin-ARPES experiments. **Funding:** This work was supported by Research Grant Council (RGC) via the early career scheme (ECS) with grant number 21304023, the Research Grant Council (RGC) via the Collaborative Research Fund (grant number C6033-22G), the Collaborative Research Equipment Grant (grant number C1018-22E), the National Natural Science Foundation of

China (12104379), and Guangdong Basic and Applied Basic Research Foundation (2021B1515130007). C.F. and D.C. thank the European Research Council (ERC Advanced Grant No.742068 `TOPMAT'), the DFG through SFB 1143 (project ID. 247310070) and the Wurzburg-Dresden Cluster of Excellence on Complexity and Topology in Quantum Matter ct.qmat (EXC2147, project ID. 390858490) for support. The SX-ARPES measurement was performed under the approval of BL25SU at SPring-8 (Proposal No.2022A2060, 2022B2106 and 2023B1691). **Author contributions:** JZM supervised this project; ZHL, WLL, XL, SYF and JZM performed ARPES experiments with the help of KY; JO, ML and BT; DC and CF synthesized the single crystals; ZHL performed first-principles calculations of the band structure; JZM and ZHL analyzed the data and plotted the figures; JWL contributed to the discussion of this project; JZM and ZHL wrote the manuscript.

**Competing interests**

The authors declare that they have no competing interests.

**Data and materials availability:** All data needed to evaluate the conclusions in the paper are present in the paper and/or the Supplementary Materials. Materials and additional data related to this paper may be requested from the authors.

**Supplementary Materials**

Method

Supplementary Text Sections 1-5.

S1. Determining photon energies versus out-of-plane $k_z$.

S2. The nature of a flat band on the Fermi level.

S3. SOC and long-range magnetic order.

S4. Bulk band splitting in soft X-ray data.

S5. EDC fitting of spin resolved data.

Supplementary Figures 1 to 8.

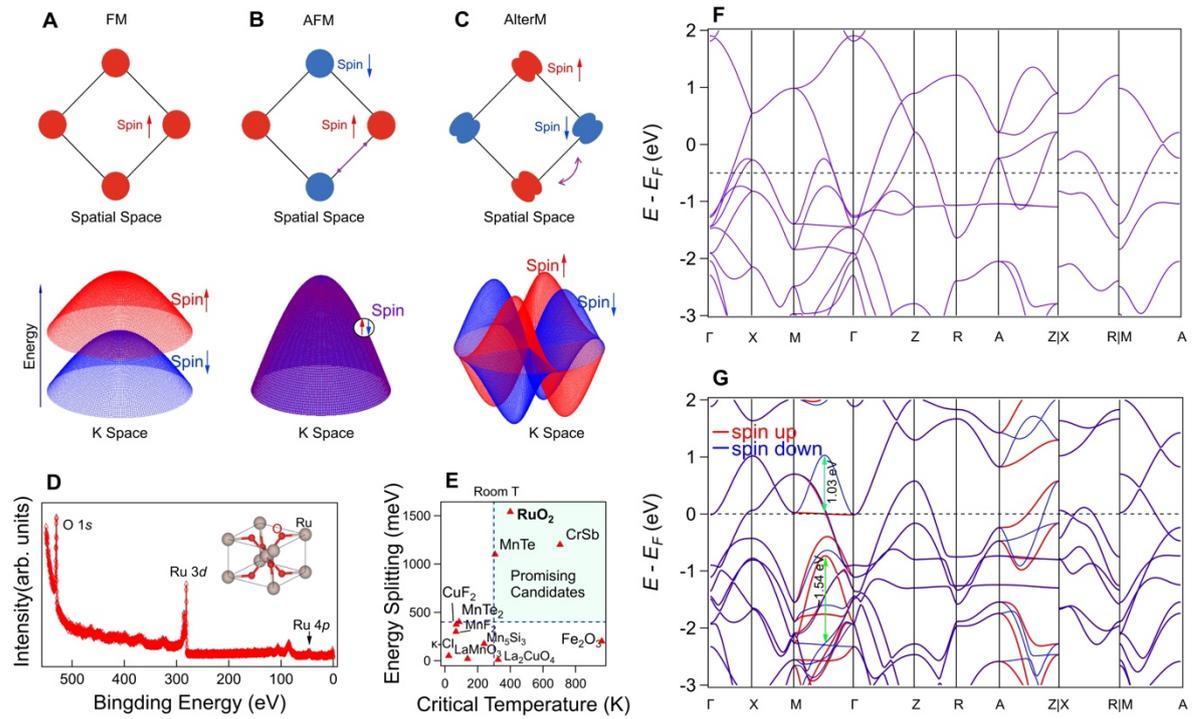

Fig. 1. Overview of d-wave altermagnetism in RuO$_2$. (A-C) Schematic of the real-space magnetic lattice, sublattices, and spin structure in reciprocal space. (D) Core levels of single crystal of RuO$_2$ displaying O 1s, Ru 3d, and 4p peaks. (E) Chart illustrating potential altermagnet candidates plotted against critical temperature and energy splitting, with the light green region indicating promising candidates for applications. This panel is adapted from *(3)* and the related references. (F, G) Electronic band structure calculations along high symmetry lines for the paramagnetic (PM) phase and altermagnetic phase of RuO$_2$, respectively. The calculated bands in (G) are shifted upward by 0.36 eV.

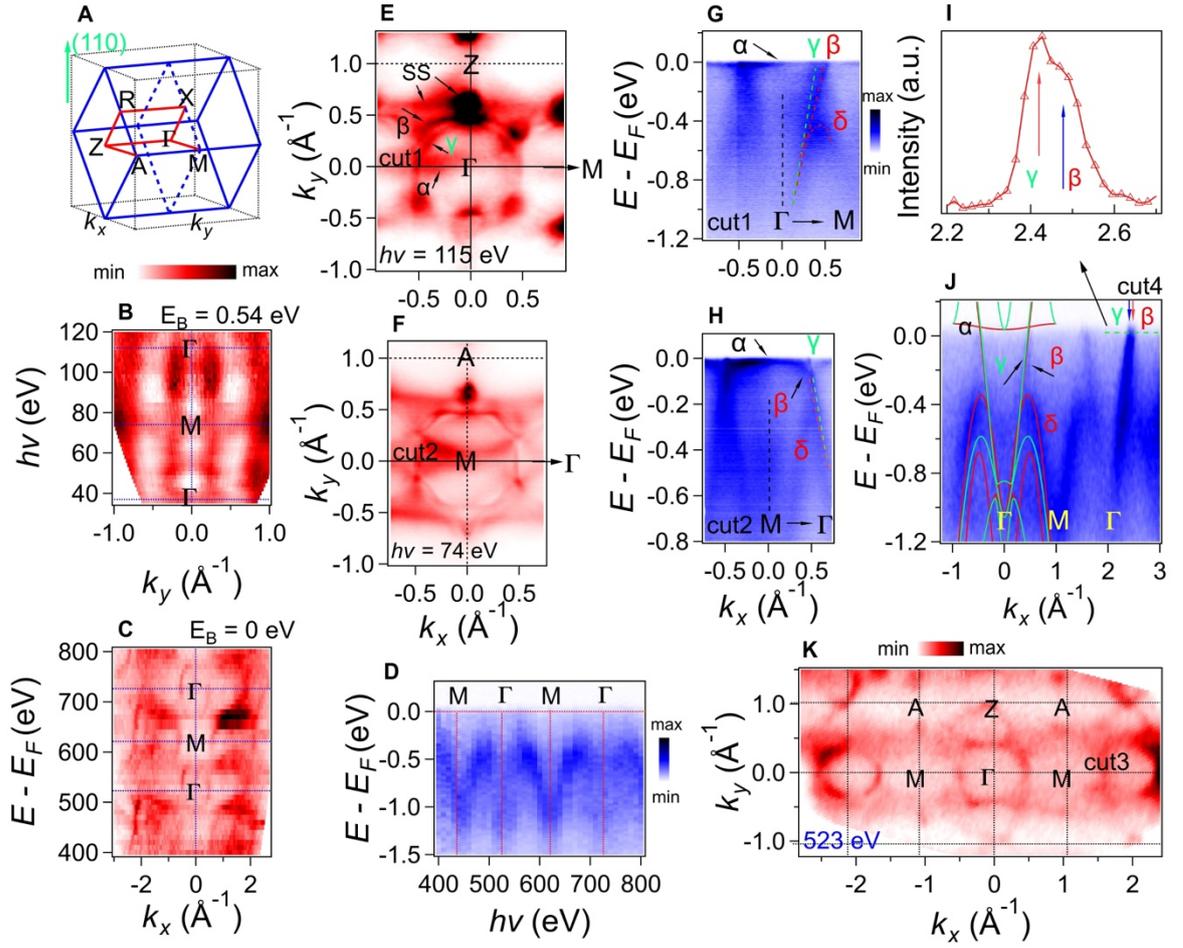

Fig. 2. Three-dimensional band structure of RuO$_2$. (A) Three-dimensional BZ (blue line) of RuO$_2$. The red lines show the high symmetry directions. The black dotted box is only a reference to help show the (110) cleavage direction. (B) Photon energy-dependent constant energy surface along the $k_y$ direction at a binding energy of 0.54 eV, obtained with photon energy ranging from 36 to 120 eV. (C) Photon energy-dependent Fermi surface along the $k_x$ direction, acquired with photon energy ranging from 400 to 800 eV. (D) Vertical cut at the BZ center of (C), highlighting the strong 3D bulk bands. (E, F) Fermi surfaces acquired with photon energies of 115 eV (Γ plane) and 74 eV (M plane), respectively. SS denotes surface states. (G, H,) ARPES-measured band structure along cut1, and cut2, respectively. (I) Momentum distribution curve on the Fermi level near the Fermi vector of β, γ bands as shown in J, indicating the separation of β and γ bands. (J) ARPES spectra along cut3. Calculated bands are provided for reference. The calculated bands are shifted upward by 0.41 eV to compare with soft X-ray ARPES data. (K) Fermi surfaces acquired with photon

energies of 523 eV (centered at Γ).

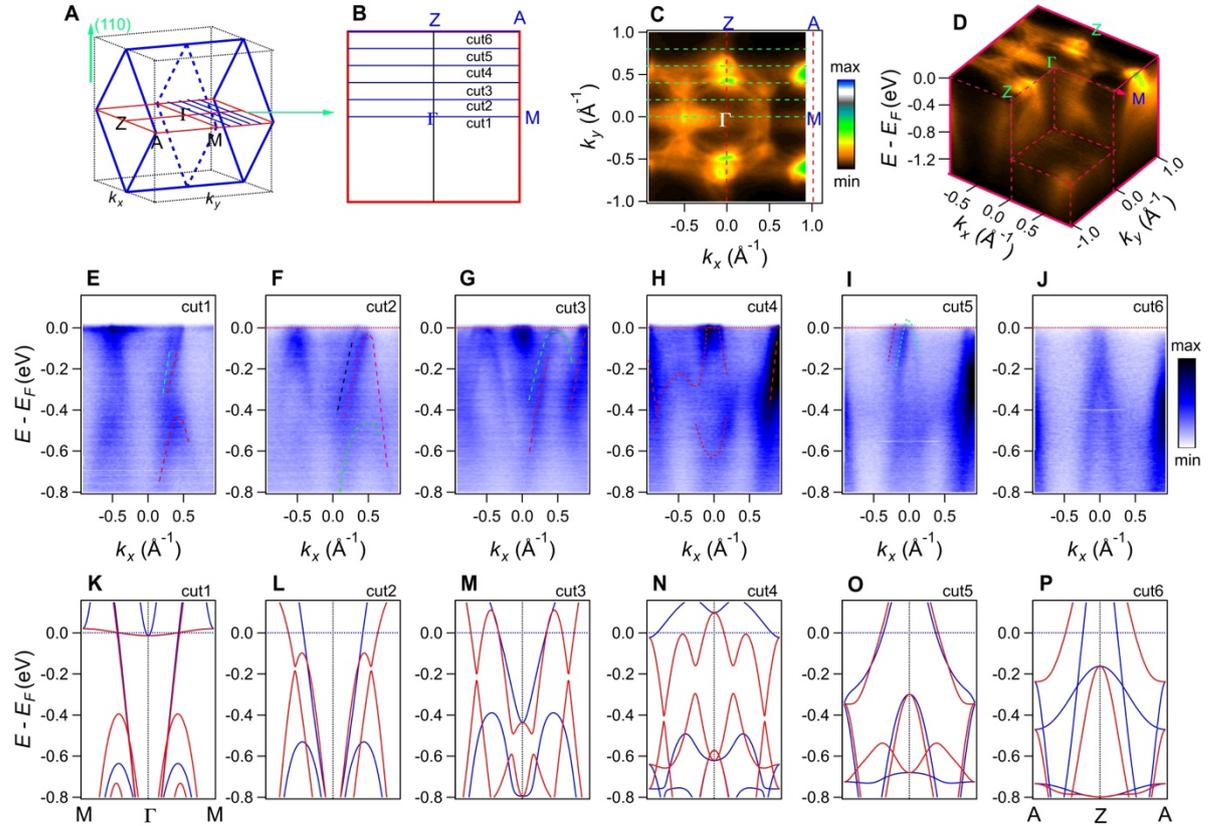

Fig.3. Band Splitting in ΓMAZ Plane. (A, B) Three-dimensional BZ of RuO$_2$ and the two-dimensional cross-section in the ΓMAZ plane. (C, D) Fermi surfaces and three-dimensional electronic band structure in the ΓMAZ plane. (E-J) ARPES spectra of cuts parallel to the ΓM direction with an equal $k_z$ offset of 0.2(ΓZ) between each. (K-P) Calculated band structure along cut1-6. All the calculated bands are shifted up by 0.36 eV to compare with UV ARPES data.

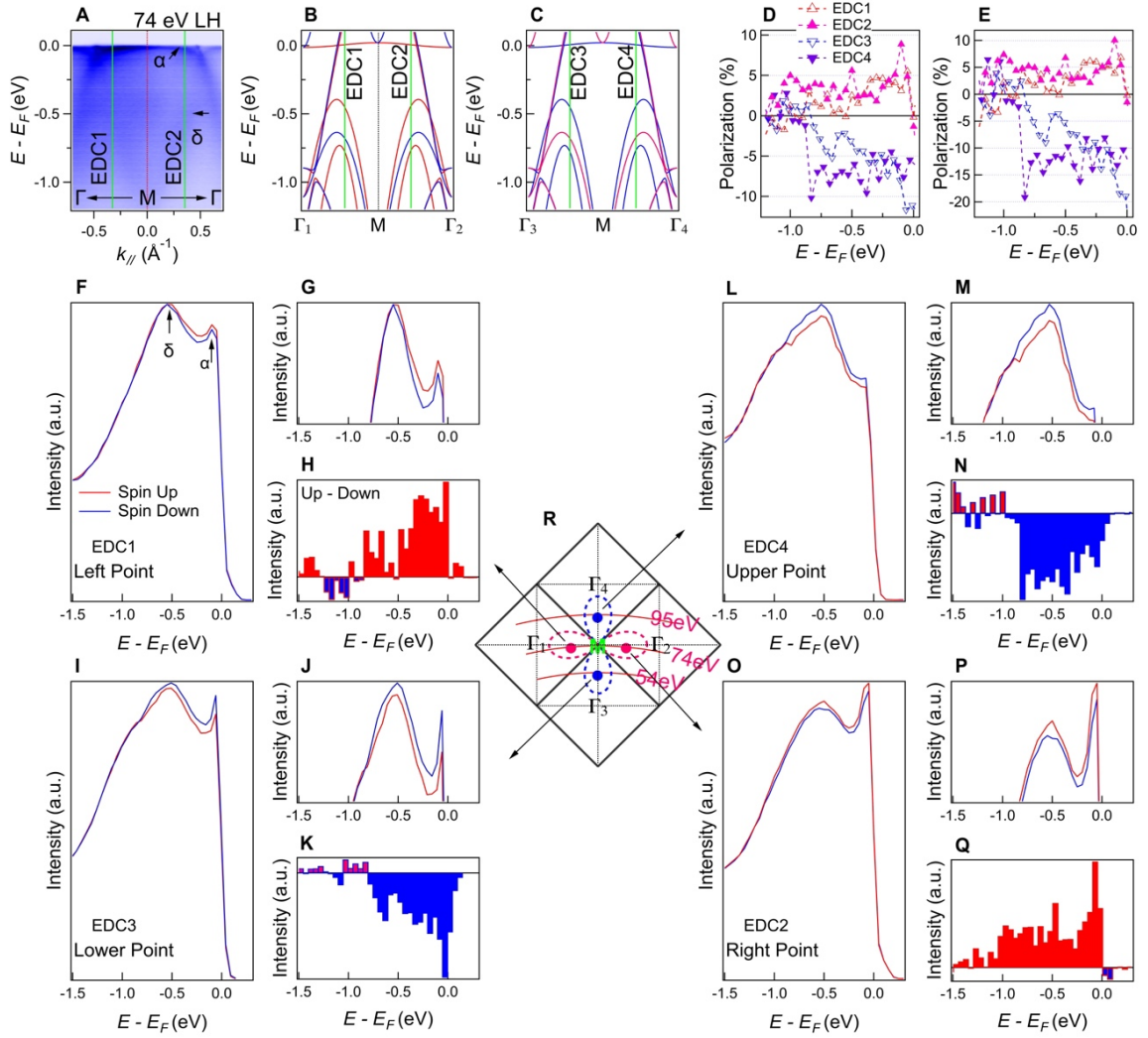

Fig. 4. Observation of d-wave spin texture in RuO2. (A) ARPES spectra centered at the M point along the ΓM direction acquired with $hv$ = 74 eV. (B, C) Calculated spin-polarized band structure along the $\Gamma_1 M \Gamma_2$ direction and $\Gamma_3 M \Gamma_4$ direction, which are perpendicular to each other, demonstrating inverted spin polarization between the perpendicular cuts. (D, E) Experimental spin polarization along EDC 1-4 symmetrically around the M point. The data in E are obtained after removing the linear background. (F, G) Spin-polarized EDCs along EDC1 at the left point and its zoomed-in view. The red curve denotes spin-up, and the blue one denotes spin-down. (H) Spin-polarized intensity difference between spin-up and spin-down, with the red color indicating spin-up polarization for α and δ bands. (I-K) Similar to (F-H) but recorded at the lower point with $hv$ = 54 eV. The blue color indicates spin-down polarization for α and δ bands. (L-N) Similar to (F-H) but recorded at the upper point with $hv$ = 95 eV. The blue color indicates spin-down polarization for α and δ bands. (O-Q) Similar

to (F-H) but recorded at the right point. The red color indicates spin-up polarization for α and δ bands. (R) The cross-section of the Brillouin Zone (BZ) and the high symmetry points, as well as the four points where the spin EDCs are acquired.

# Supplementary Materials For
# Observation of Giant Spin Splitting and d-wave Spin Texture in Room Temperature Altermagnet $RuO_2$

Zihan Lin[1,†], Dong Chen[2,3,†], Wenlong Lu[1,†], Xin Liang[1], Shiyu Feng[1], Kohei Yamagami[4], Jacek Osiecki[5], Mats Leandersson[5], Balasubramanian Thiagarajan[5], Junwei Liu[6], Claudia Felser[3], Junzhang Ma[1,7,*]

[1]*Department of Physics, City University of Hong Kong, Kowloon, Hong Kong, China*
[2]*College of Physics, Qingdao University, Qingdao 266071, China*
[3]*Max Planck Institute for Chemical Physics of Solids, 01187 Dresden, Germany*
[4]*Materials Sciences Research Center, Japan Atomic Energy Agency, Sayo, Hyogo 679-5148, Japan*
[5]*MAX IV laboratory, Fotongatan 8, 22484 Lund, Sweden*
[6]*Department of Physics, The Hong Kong University of Science and Technology, Hong Kong, China.*
[7]*City University of Hong Kong Shenzhen Research Institute, Shenzhen, China*

*†The authors contributed equally to this work.*
*\*Corresponding to: Junzhang Ma (junzhama@cityu.edu.hk)*

These Supplementary Materials contain:

**Method**

**Supplementary Sections 1-5.**

S1. Determining photon energies versus out-of-plane $k_z$.

S2. The nature of a flat band on the Fermi level.

S3. SOC and long-range magnetic order.

S4. Bulk band splitting in soft X-ray data.

S5. EDC fitting of spin resolved data.

**Supplementary Figures 1 to 8.**

**Method**

Single crystals of $RuO_2$ were grown by the chemical vapor transport method. $RuO_2$ powder together with the transport agent $TeCl_4$ (5 mg/cm$^3$) was sealed in an evacuated quartz tube. The quartz tube was then placed into a two-zone tube furnace with 1100 °C at the raw material side and 1000 °C at the growth side. The reaction was maintained for one week, and millimeter-sized crystals were obtained.

Conventional ARPES measurements were performed at Bloch beam line of MAX-IV synchrotron, the energy and angular resolutions were set to ~20 meV and 0.1°, respectively, and the temperature was set to around 22K; Soft X-ray ARPES measurements were performed at BL25SU beamline of SPring-8 Synchrotron and the energy and angular resolutions were set to ~60 meV and 0.1°, and the temperature was set to around 77K *(56,57)*, respectively; Spin-ARPES measurements were performed at DREAMLINE beamline of Shanghai Synchrotron Radiation Facility (SSRF) and the energy resolution were set to ~100 meV, and the temperature was set to around 11K. The one-direction spin detector measures spin direction parallel to the sample surface aligning with ΓM direction during the measurements. The samples for all ARPES measurements were cleaved in situ and measured in a vacuum better than 2×10$^{-10}$ Torr.

We have calculated the ground state electronic structure of $RuO_2$. All DFT calculations were performed within the Perdew-Burke-Ernzernhof (PBE) *(58)* generalized gradient approximation and projector augmented wave (PAW) method *(59)* using QUANTUM ESPRESSO package *(60,61)* with pseudopotential from the PSLibrary *(62)*. The energy cutoff for the plane wave basis is 80 Ry with a charge density cutoff of 400 Ry. A Monkhorst-Pack *(63)* 12×12×18 *k*-mesh has been used. The electronic correlations on Ru atoms were accounted for using the DFT+U *(64)* with U$_{eff}$ = 3 eV. The Wannier functions have been built up using WANNIER90 code *(65)*. The surface states have been calculated with WannierTools *(66)*.

## S1. Determining photon energies versus out-of-plane $k_z$

In the main text, we used UV light for the spin texture validation due to its high energy resolution and photon intensity. Low photon energy, however, causes a stronger $k_z$ broadening effect and strengthens surface states, making the determination of out-of-plane $k_z$ relatively difficult compared to soft X-ray data. Accurately determining the momentum in 3D reciprocal space is important for d-wave altermagnets in $RuO_2$, since the spin texture represents bulk information. To obtain accurate determination of the right photon energy and locations in momentum space, both UV and soft X-ray photon energy-dependent mappings were carried out, as shown in Fig. 2B and C in the main text. An important feature for verifying the accuracy of $k_z$ is the central point which alters with photon energy between Γ and M. Photon energies versus $k_z$ were easily determined with the use of soft X-ray ARPES data in Fig. 2C, which clearly reveals a clear periodic structure as illustrated in Fig. S1C. The periodic structure is in good alignment with the lattice parameter along the out-of-plane direction, which allows us to identify the Γ plane at 521 eV, the M plane at 621 eV, and the Γ plane again at 725 eV.

On the other hand, the out-of-plane dispersion in UV ARPES data is weaker but still noticeable as shown in Fig. 2B, where we identified 37 eV as the Γ plane, 74 eV as the M Plane, and 115 eV as the Γ plane once more. This determination is consistent with the extension of the $hv$-$k_z$ curve from the soft X-ray region to the UV region, as seen in Fig. S1C. Additionally, the bands that cross the BZ center should cycle between MΓM and ΓMΓ, providing compelling evidence that the $k_z$ determination is accurate. In Fig. S1 D-H, we plotted the ARPES spectra crossing the BZ center at photon energies of 37, 74, 115, 521, and 621 eV. As expected, the slopes' sign of the (β, γ) bands alternates between positive and negative, as denoted with green dotted lines in Fig. S1 D-H. Therefore, the determination of the 3D BZ band the bulk band nature is guaranteed by the combination of soft X-ray and UV ARPES data.

## S2. The nature of α flat band on the Fermi level and surface states

The bulk band character of the β, γ and δ bands can be confirmed by the band structure altering between MΓM and ΓMΓ at different photon energies, in previous discussions.

Nevertheless, the α flat band on the Fermi level is not visible in the soft X-ray data. Fig. S2 provides the results of a surface band structure projection calculation to examine if the α band indicates a surface state. The calculations reveal no discernible evidence of surface bands near the Fermi level along the ΓM direction. Instead, the broad projection of the α band, indicated by the dotted ellipse, demonstrates its bulk nature.

In an earlier study, a drum-head surface state is claimed to result from the bulk nodal line, on the Fermi level at the same position of the α flat band *(55)*. Our computational investigation, however, does not reveal any unique surface band separated from bulk bands at the Fermi level. This implies that such surface states, if they exist, are topologically unprotected and overlap with the bulk band projection. They can be viewed as a top layer extension of the wavefunction of the same bulk flat band. This observation suggests that the α band mainly consists of bulk spectra. The disappearance of the α band in the soft X-ray data is attributed to a little chemical potential shift between the surface and bulk area caused by the band bending effect. This phenomenon is commonly observed in other materials, such as the conduction bulk band in the topological insulator $Bi_2Se_3$ *(67)*.

On the other hand, our calculation does show surface states along the ΓM direction, as indicated by SS in Fig. S2B. The lower surface band is located at around -0.25 eV which is above the δ band. Our experimental data also show the presence of such a surface band. Fig. S3 plots the cuts with three different photon energies $hv$ = 54 eV, 74 eV and 94 eV and with three polarizations, i.e., circular plus (CP), linear horizontal (LH), and linear vertical (LV), respectively. The α-δ bands are observed within LH data and show good agreement with the calculations. On the other hand, in the LV data, one additional band is significantly apparent above δ band located at around -0.25 eV. The position and dispersion agree well with the surface band predicted in Fig. S2B. According to the experimental findings, surface bands are strengthened under LV polarization while bulk bands are apparent under LH polarization. Circular polarization, on the other hand, has both surface and bulk states.

**S3. Spin-Orbit Coupling (SOC) and long-range magnetic order**

In the main text, we highlighted the difference between band calculations of the paramagnetic and altermagnetic phases, arguing that band splitting provides strong evidence of long-range magnetic order. Meanwhile, a relevant question arises: Is it possible for SOC to be responsible for band splitting in a non-magnetic order instead of just being caused by magnetic order? To address this concern, we carried out band structure calculations in the paramagnetic phase with SOC, as shown in Fig. S4. Six cuts in the ΓMAZ plane were plotted, revealing no band splitting in all calculations.

On the other hand, band structure calculations in the altermagnetic phase with SOC, as shown in Fig. S4 C, revealed that SOC does not introduce additional bands but instead enhanced band hybridization, as indicated by the black ellipse. Fig. 4D shows the same cut without SOC for comparison. Upon a thorough analysis of the experimental data, we observed a similar hybridization between the α and β bands in the ARPES data, as shown in Fig. S5 A and B. It should be noted that SOC was not considered in the calculations shown in the main text, because doing so would have rendered spin an invalid good quantum number.

In summary, the band splitting observed in our experiments provides unambiguous proof of the long-range magnetic order in $RuO_2$.

## S4. Investigation of Bulk band splitting in soft X-ray data

We provided a thorough examination of the band structure in the ΓMAZ plane in the main text, demonstrating band spin splitting that is in good agreement with first principle calculations in the altermagnetic phase. Analogous studies were carried out inside the soft X-ray spectrum. We focused on data acquired at 521 eV photon energy, concentrating on band structure within the same plane. Fig. S7 A-C shows the relevant BZ and Fermi surface. A set of bands labelled as cut 1-6 that run parallel to the direction of the ΓM are shown for the experimental data in Fig. S7 D-I and for the calculations in Fig. 7 J-O.

Despite the lower energy resolution and broader bands in the soft X-ray experimental data, the general band dispersion remains consistent with that acquired using UV light. The band splitting predicted by the calculations is visible in cut 1-4, as highlighted in the data. This comprehensive comparison provides robust evidence supporting the agreement between the experimental observations and theoretical predictions under the altermagnetic phase.

## S5. EDC fitting of spin resolved data

One notable feature of the spin EDCs shown in Fig. 4 of the main text is the prominent background, which could cause the spin polarization to be underestimated. To remove this negative impact, we applied a fitting process for each of the four EDCs, which included a simple linear background and Gaussian peaks multiplied by a Fermi-Dirac function (Fig. S6). Fig. S8 A, C, E, and G show the fitting results. The main goal of the fitting is to extract the linear background. Black dotted lines represent the fitted backgrounds (multiplied by Fermi-Dirac function). Interestingly, both the upper and lower points show a broad peak at approximately -1 eV, whereas the left and right points show no such peak. As the calculation predicted, these peaks are attributed to the valence bands below the $\delta$ band. These bands are too weak to be seen in the left and right points due to matrix element effects. An equal linear background was subtracted for both spin-up and spin-down EDCs at the same position of the four points to remove the influence of the background on the spin difference between the spin up and down of EDCs. As seen by the blue curves in Fig. S8 B, D, F, and H, this background reduction greatly increased the spin polarization, with the maximum reaching up to 20% for both the upper and lower points.

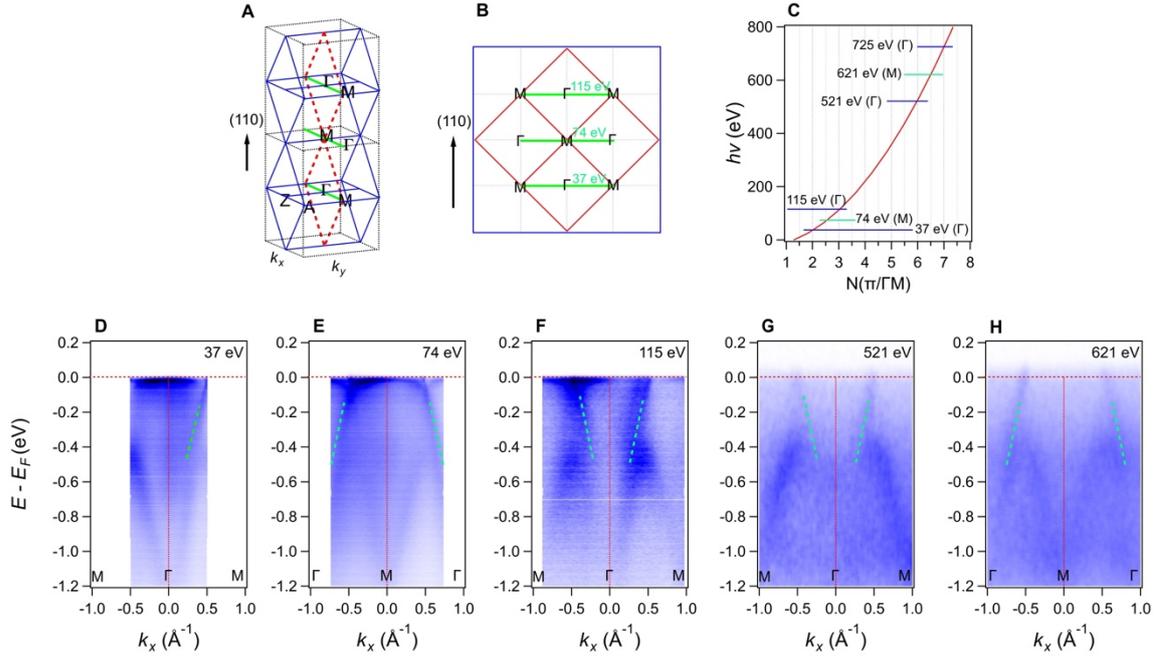

Fig. S1. Relationship between photon energy and out-of-plane $k_z$. (A) Schematic of the 3D BZs stacked along the out-of-plane direction. The green lines represent cuts along $k_x$, alternating versus photon energy between ΓM and MΓ. (B) Cross-section of the Brillouin zone along the out-of-plane direction. (C) The relationship between photon energy and $k_z$ for $RuO_2$ obtained from the photon energy-dependent data, as shown in Fig. 2 B, C in the main text. (D-H) ARPES spectra along $k_x$ under photon energies of 37 eV, 74 eV, 115 eV, 521 eV, and 621 eV, respectively.

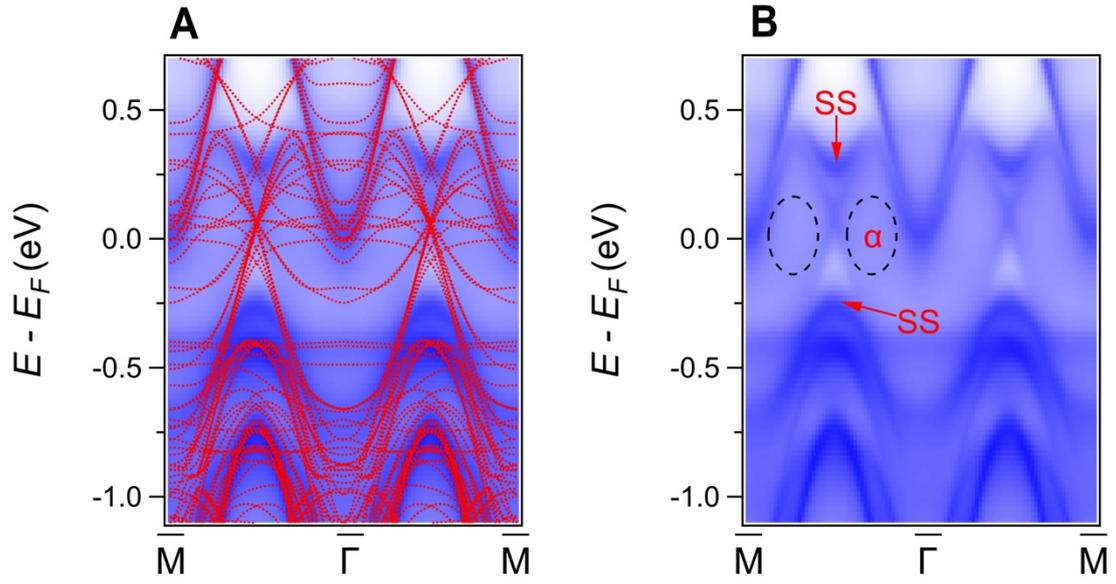

Fig. S2. (110) Direction projected surface states calculations including SOC. (A) Surface-projected band structure with bulk bands (11 $k_z$ evenly distributed between $k_z = 0$ and $\sqrt{2}\pi/a$) plotted alongside for comparison. (B) Pure surface-projected band structure. The black dotted area represents the projection of the bulk α band without any sharp surface signature.

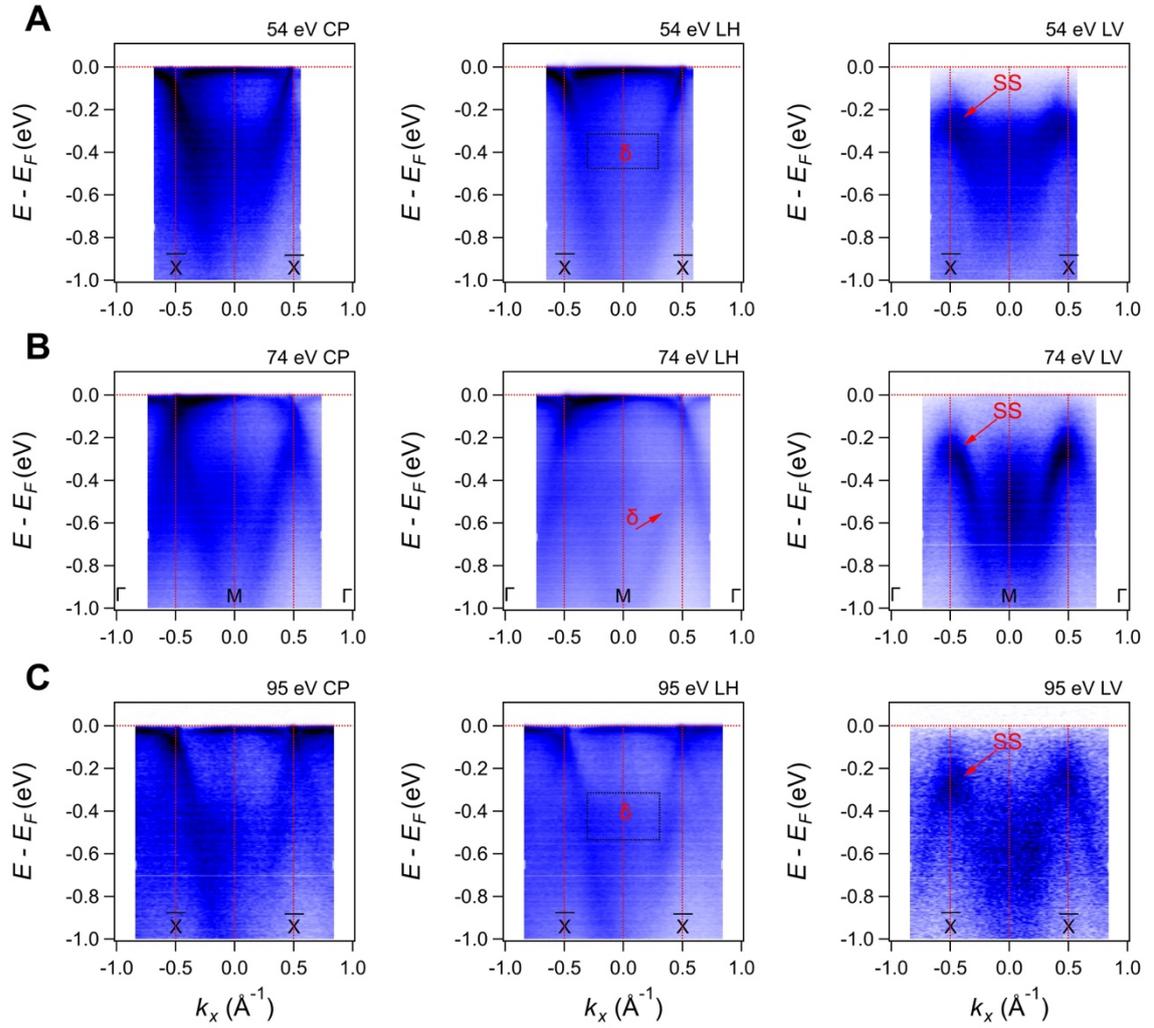

Fig. S3. Light polarization dependent cuts at 54 eV (A), 74 eV (B) and 95 eV (C) with circular plus (CP), linear horizontal (LH) and linear vertical (LV) photons, respectively. SS denotes surface states.

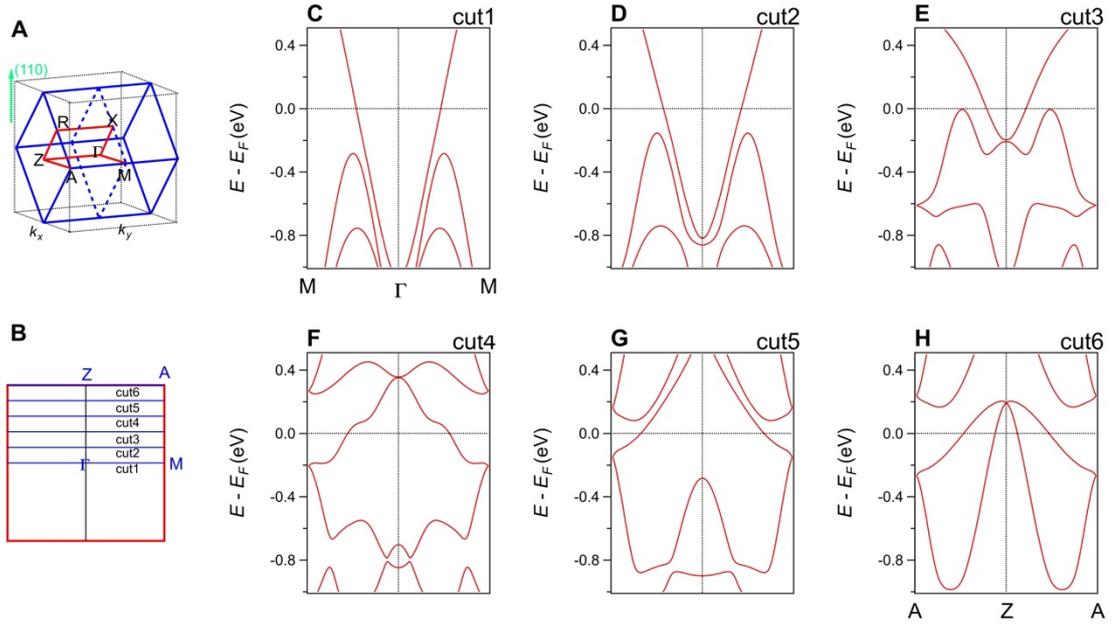

Fig. S4. Band structure calculations for non-magnetic RuO$_2$, indicating no band splitting in the whole BZ. (A) Schematic diagram of the BZ. (B) Cross-section of the BZ along the $k_x - k_y$ plane. (C-H) Non-magnetic DFT-calculated bands including SOC for cut1 to cut6, respectively. No band splitting is observed.

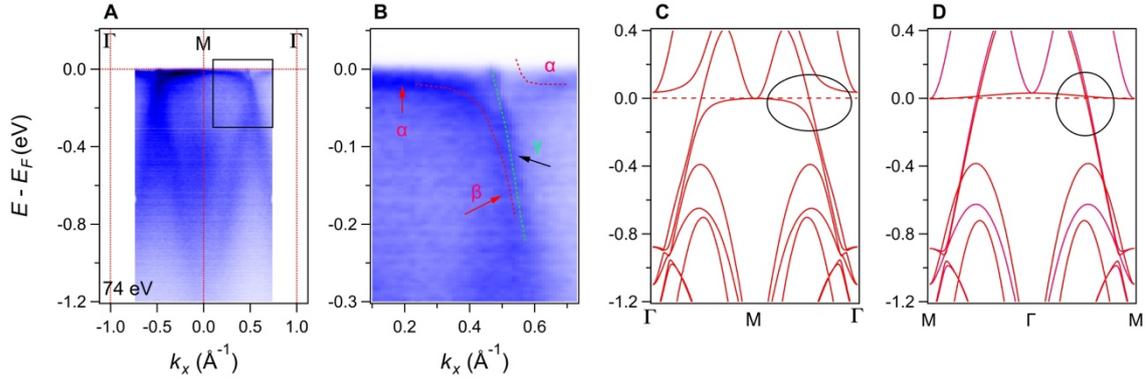

Fig. S5. SOC-induced band hybridization gap. (A, B) ARPES spectra and its zoomed-in view along ΓMΓ direction acquired with $hv$ = 74 eV. (C) Calculated band structure along ΓMΓ direction under the altermagnetic phase considering SOC. (D) Same as (C) but without considering SOC.

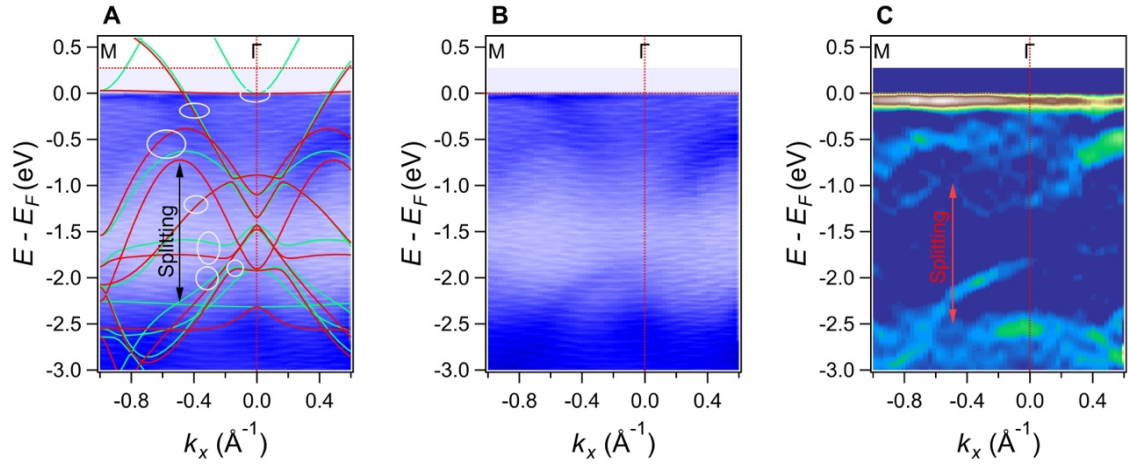

Fig. S6. Giant spin splitting of the up occupied bands along ΓM direction. (A) APRES spectra with a wide energy range along ΓM direction. The calculated band structure is overlaid for comparison. The spin splitting is indicated by the double arrow. The gray ellipses mark the Kramer pairs. (B) Pure ARPES spectra same as in B. (C) Curvature intensity plot of the experimental data. The double arrow points to the signature of the band splitting.

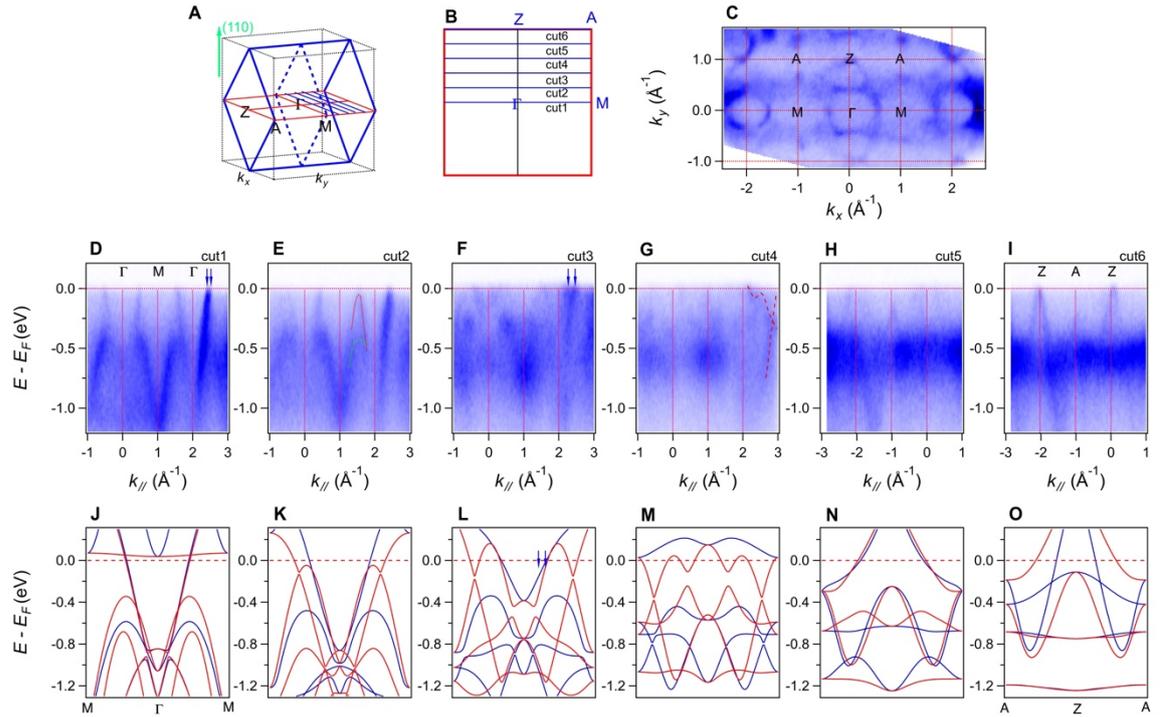

Fig. S7. Soft X-ray ARPES Data in ΓMAZ Plane. (A, B) Three-dimensional BZ of RuO$_2$ and the two-dimensional cross-section in the ΓMAZ plane. (C) Fermi surfaces in the ΓMAZ plane acquired with $hv$ = 521 eV. (D-I) ARPES spectra of cuts parallel to the ΓM direction with an equal $k_z$ offset of 0.2*ΓZ between each. (J-P) Calculated band structure along cut1-6 under the altermagnetic phase. The red color denotes spin up, and the blue color denotes spin down. The bands are shifted up by 0.41 eV for comparison with soft X-ray ARPES data.

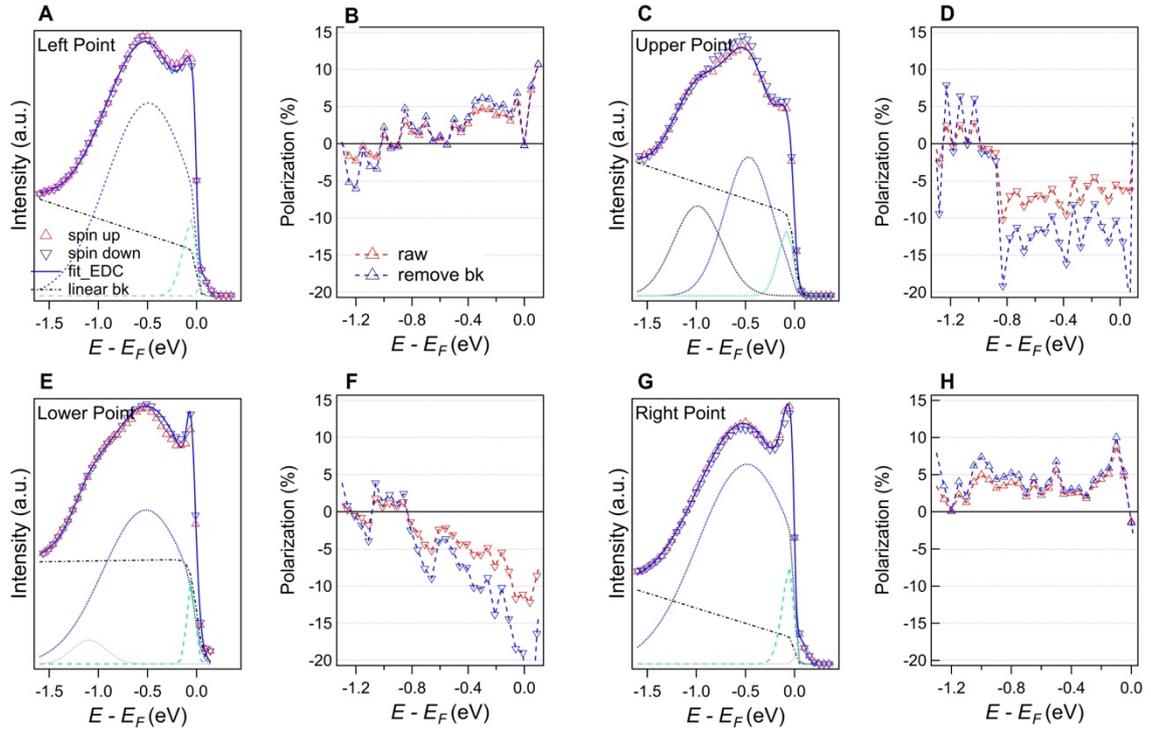

Fig. S8. Spin-polarized EDCs and polarization after removing linear background. (A) Spin-resolved EDC of the left point, where the triangle point represents the raw data, dashed line represents the linear background, and the solid line represents the fitted EDC. The blue and green dotted lines represent the fitted peaks. (B) The polarization EDCs. The red dotted curve denotes the pristine polarization, and the blue dotted curve denotes the polarization derived after removing the linear background. (C-D, E-F, G-H) Same as A-B, but for the upper point, lower point, and right point, respectively.